\documentclass{article}
\usepackage{spconf,amsmath}

\usepackage[utf8]{inputenc} 
\usepackage[T1]{fontenc}    
\usepackage{hyperref}       
\usepackage{url}            
\usepackage{booktabs}       
\usepackage{amsfonts}       
\usepackage{nicefrac}       
\usepackage{microtype}      

\usepackage{subcaption}
\usepackage{multicol}
\usepackage{algorithm}
\usepackage{epsfig}
\usepackage{graphicx}
\usepackage{amsmath}
\usepackage{amssymb}

\title{Learned Variable-Rate Image Compression with Residual Divisive Normalization}

\name
{Mohammad Akbari$^*$, Jie Liang$^*$, Jingning Han$^\dagger$, Chengjie Tu$^\ddagger$}
\address{akbari@sfu.ca, jiel@sfu.ca, jingning@google.com, chengjietu@tencent.com\\Simon Fraser University, Canada$^*$, Google Inc.$^\dagger$, Tencent Technologies$^\ddagger$}

\begin{document}\sloppy

\def\x{{\mathbf x}}
\def\L{{\cal L}}

%

\maketitle

\begin{abstract}
Recently it has been shown that deep learning-based image compression has shown the potential to outperform traditional codecs. However, most existing methods train multiple networks for multiple bit rates, which increases the implementation complexity. In this paper, we propose a variable-rate image compression framework, which employs more Generalized Divisive Normalization (GDN) layers than previous GDN-based methods. Novel GDN-based residual sub-networks are also developed in the encoder and decoder networks. Our scheme also uses a stochastic rounding-based scalable quantization. To further improve the performance, we encode the residual between the input and the reconstructed image from the decoder network as an enhancement layer. To enable a single model to operate with different bit rates and to learn multi-rate image features, a new objective function is introduced. Experimental results show that the proposed framework trained with variable-rate objective function outperforms all standard codecs such as H.265/HEVC-based BPG and state-of-the-art learning-based variable-rate methods.
\end{abstract}

\section{Introduction}
\label{Introduction}

In the last few years, deep learning-based image compression \cite{agustsson2017soft,akbari2019dsslic,balle2016,minnen2018joint,theis2017lossy,toderici2015variable,toderici2017full} has made tremendous progresses, and has achieved better performance than JPEG2000 and the H.265/HEVC-based BPG image codec \cite{bellard2017bpg}. 

In \cite{balle2016}, a generalized divisive normalization (GDN)-based scheme was proposed. The encoding network consists of three stages of convolution, subsampling, and GDN layers. Despite its simple architecture, it outperforms JPEG2000 in both PSNR and SSIM. A compressive autoencoder (AE) framework with residual connection as in ResNet was proposed in \cite{theis2017lossy}, where the quantization was replaced by a smooth approximation, and a scaling approach was used to get different rates. In \cite{agustsson2017soft}, a soft-to-hard vector quantization approach was introduced, and a unified framework was developed for both image compression and neural network model compression. 

In \cite{akbari2019dsslic}, a deep semantic segmentation-based layered image compression (DSSLIC) was proposed, by taking advantage of the Generative Adversarial Network (GAN). A low-dimensional representation and segmentation map of the input were encoded. Moreover, the residual between the input and the synthesized image was also encoded. It outperforms the BPG codec (in RGB444 format) in both PSNR and MS-SSIM \cite{wang2003multiscale} across a large range of bit rates.

In \cite{minnen2018joint}, a context model of entropy coding for end-to-end optimized image compression was proposed, and a hyper-prior was augmented with the context. This method represents the state-of-the-art learning-based image compression, which outperforms BPG in terms of both PSNR and MS-SSIM. Another context-based model was proposed in \cite{li2019learning}, where an importance map for locally adaptive bit rate allocation was employed to handle the spatial variation of image content.

All the above-mentioned methods train multiple networks for multiple bit rates, which increases the implementation complexity. Therefore the variable-rate approach is desired in many scenarios, in which a single neural network model is trained to operate at multiple bit rates with satisfactory performance \cite{toderici2015variable,toderici2017full,cai2018efficient,zhang2018learned}.

In \cite{toderici2015variable}, long short-term memory (LSTM)-based recurrent neural networks (RNNs) and residual-based layer coding was used to compress thumbnail images. Better SSIM results than JPEG and WebP were reported. This approach was generalized in \cite{toderici2017full}, which proposed a variable-rate framework for full-resolution images by introducing a gated recurrent unit, residual scaling, and deep learning-based entropy coding. This method can outperform JPEG in terms of PSNR.

In \cite{cai2018efficient}, a CNN-based multi-scale decomposition transform was optimized for all scales. Rate allocation algorithms were also applied to determine the optimal scale of each image block. The results in \cite{cai2018efficient} were reported to be better than BPG in MS-SSIM. In \cite{zhang2018learned}, a learned progressive image compression model was proposed, in which bit-plane decomposition was adopted. Bidirectional assembling gated units were also introduced to reduce the correlation between different bit-planes \cite{zhang2018learned}.



In this paper, we propose a new deep learning-based variable-rate image compression framework, which employs more GDN layers than \cite{balle2016,minnen2018joint}. Two novel types of GDN-based residual sub-networks are also developed in the encoder and decoder networks, by incorporating the shortcut connection in ResNet \cite{he2016deep}. Our scheme uses the stochastic rounding-based scalable quantization as in \cite{gupta2015deep,Raiko15,toderici2015variable}. To further improve the performance, we encode the residual between the input and the reconstructed image from the decoder network as an enhancement layer. To enable a single model to operate with different bit rates and to learn multi-rate image features, a new variable-rate objective function is introduced that considers the performance at multiple rates. 
Experimental results show that the proposed framework trained with variable-rate objective function outperforms state-of-the-art learning-based variable-rate methods as well as all standard codecs including H.265/HEVC-based BPG (in all formats) in terms of MS-SSIM metric.

\section{The Proposed Method}
\label{Proposed Model}

\begin{figure}
 \centering
\begin{subfigure}[b]{1\linewidth}
 \centering
  \centerline{\includegraphics[width=\textwidth]{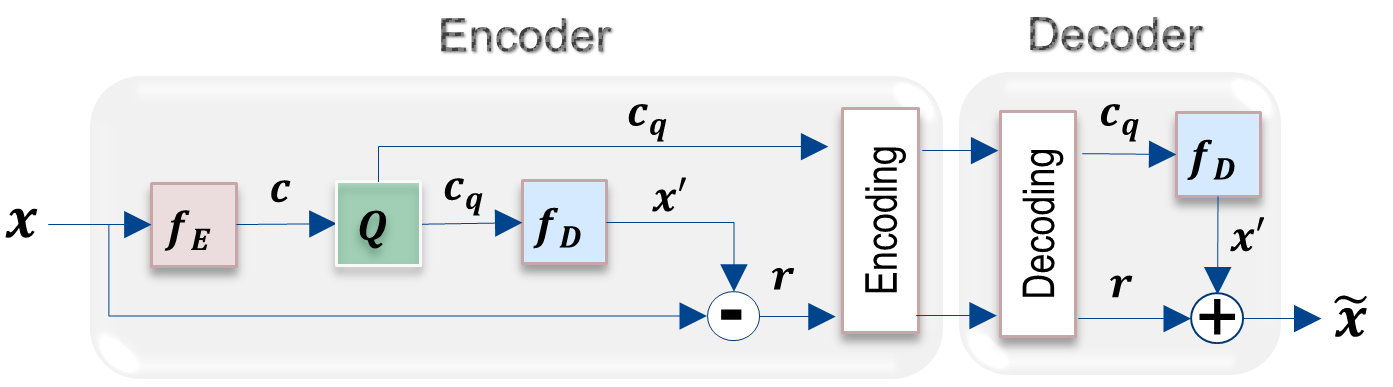}}
\end{subfigure}
\caption{Overall framework of the proposed codec ($f_E$: deep encoder, $c$: code map, $Q$: uniform scalar quantizer, $c_q$: quantized code map, $f_D$: deep decoder, $r$: residual image). }
\vskip -10pt
\label{fig:framework}
\end{figure}

The overall framework of the proposed codec is shown in Fig. \ref{fig:framework}. At the encoder side, two layers of information are encoded: the encoder network output (code map) and the residual image. The code map $c$ is a low-dimensional feature map of the original image $x$, obtained by the deep encoder $f_E$, which is quantized by a uniform scalable quantizer $Q$, and then encoded by the FLIF lossless codec \cite{sneyers2016flif}. To further improve the performance, the reconstruction of the input image (denoted by $x'$) from the quantized code map $c_q$ is obtained using the deep decoder $f_D$, and the residual $r$ between the input and the reconstruction is encoded by the BPG codec as an enhancement layer \cite{akbari2019dsslic}. At the decoder side, the reconstruction $x'$ from the deep decoder and the decoded residual image $r'$ are added to get the final reconstruction $\tilde{x}$.

It has been shown that end-to-end optimization of a model including cascades of differentiable linear-nonlinear transforms has better performance over the traditional linear ones \cite{balle2016}. One example is the GDN, which is very efficient in gaussianizing local statistics of natural images. It also provides significant improvements when utilized as a prior for image compression when used with scalable quantization. GDN was first introduced in \cite{balle2016} for a learning-based image compression framework, which had a simple architecture of 3 downsampling convolutions, each is followed by a GDN layer.

\begin{figure*}
 \centering
\begin{subfigure}[b]{0.95\linewidth}
 \centering
  \centerline{\includegraphics[width=\textwidth]{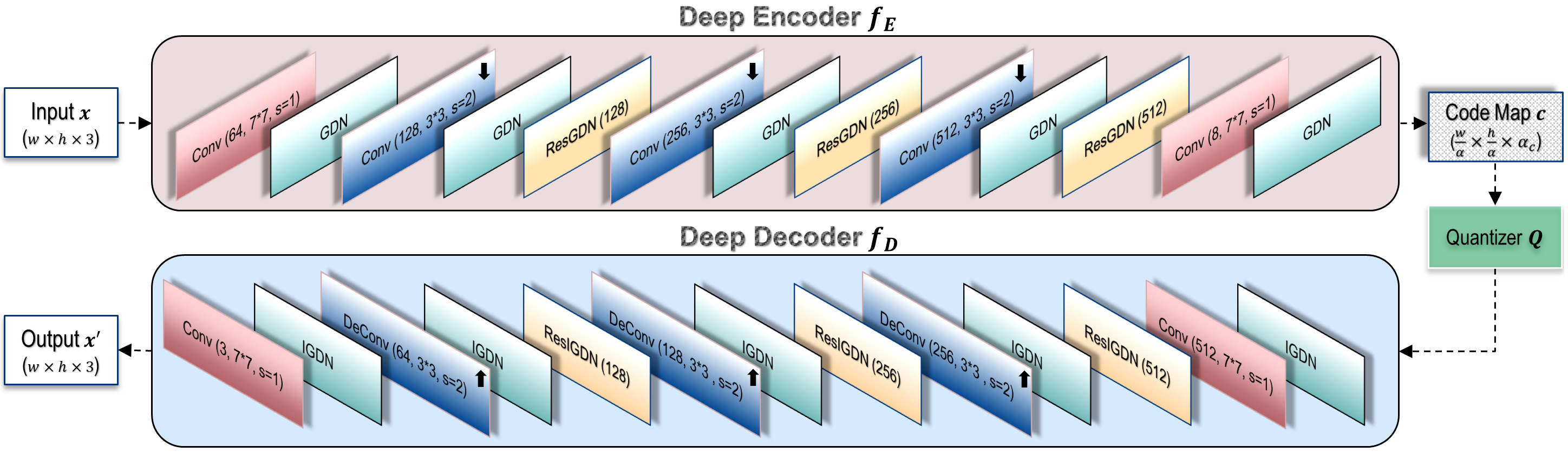}}
\end{subfigure}
\caption{Architecture of the proposed deep encoder and deep decoder networks. \textbf{Conv/DeConv (n, k$\times$k, s)}: convolution/deconvolution with n filters of size k$\times$k and stride of s. 
}
\vskip -10pt
\label{fig:deepframework}
\end{figure*}

\begin{figure}
\centering
\begin{subfigure}[b]{0.91\linewidth}
 \centering
  \centerline{\includegraphics[width=\textwidth]{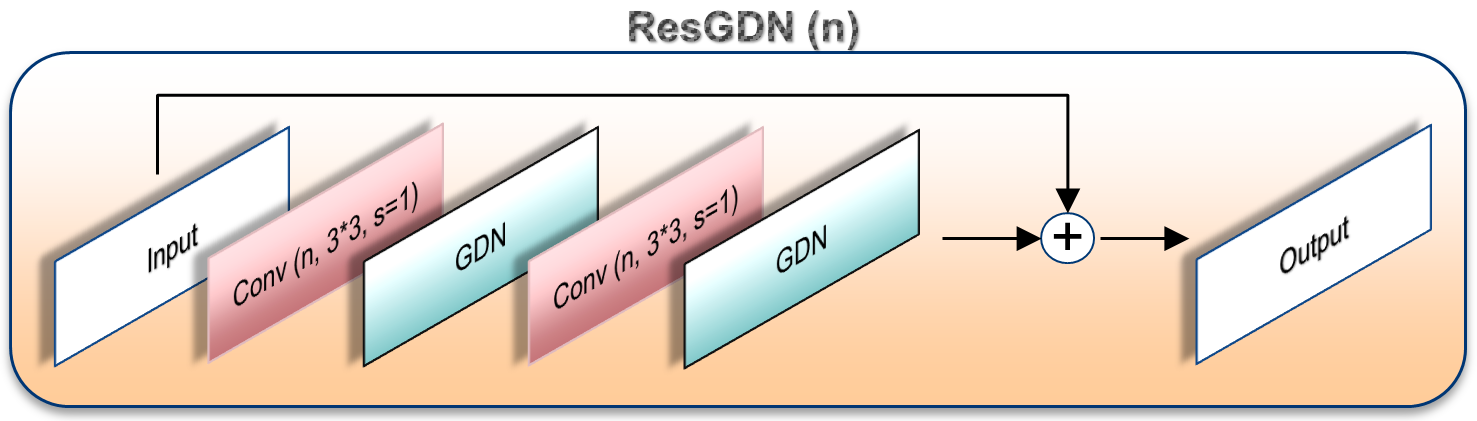}}
\end{subfigure}
\begin{subfigure}[b]{0.91\linewidth}
 \centering
  \centerline{\includegraphics[width=\textwidth]{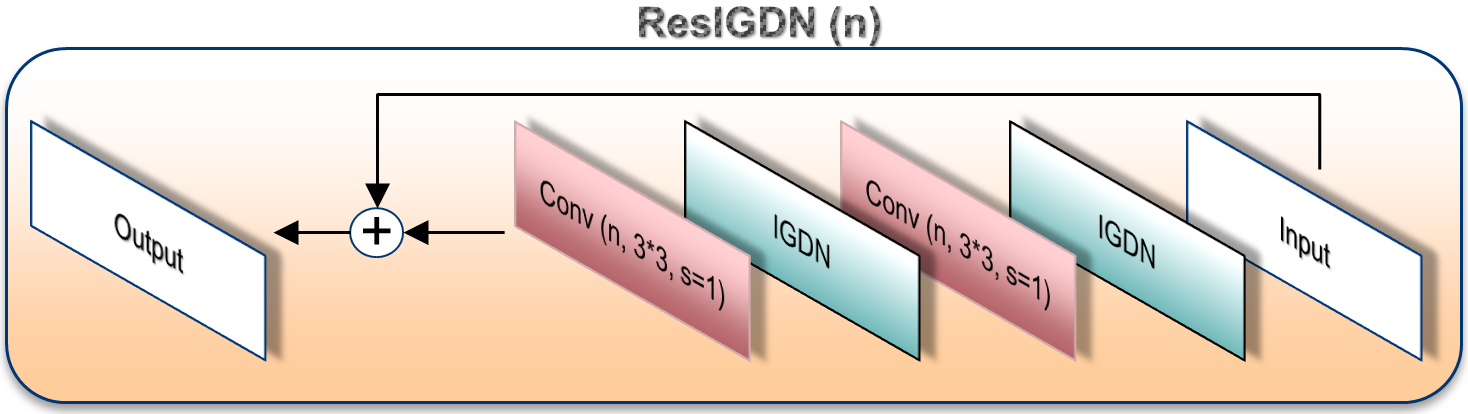}}
\end{subfigure}
\caption{Architecture of proposed ResGDN and ResIGDN transforms (n: the channel size used in the affine convolutions).}
\vskip -10pt
\label{fig:ResBlocks}
\end{figure}

The architecture of the proposed deep encoder and decoder networks are shown in Figure \ref{fig:deepframework}. Several modifications to the GDN-based schemes in \cite{balle2016,minnen2018joint} are developed. First, we adopt a deeper architecture including 11 convolution layers, followed by GDN or inverse GDN (IGDN) in the encoder and decoder. Second, for deeper learning of image statistics and faster convergence, the concept of identity shortcut connection in the ResNet in \cite{he2016deep} is introduced to some GDN and IGDN layers, denoted by ResGDN and ResIGDN. The architecture of the ResGDN and ResIGDN are shown in Figure \ref{fig:ResBlocks}. Unlike the traditional residual blocks where the ReLU and batch (or instance) normalization are employed, we utilize GDN and IGDN layers in our residual blocks, which provide better performance and faster convergence rate.

\subsection{Deep Encoder}

Let $x \in \mathbb{R}^{h\times w\times 3}$ be the input image, the code map $c \in \mathbb{R}^{\frac{h}{\alpha} \times \frac{w}{\alpha} \times \alpha_c}$ is generated using the parametric deep encoder $f_{E}$ as: $c=f_{E}(x;\Phi)$,
where $\Phi$ is the parameter vector to be optimized. The encoder consists of 5 stages where the input to the $k$th stage is denoted by $U^{(k)}$. The image $x$ is then represented as $U^{(0)}$, which is the input to the first stage of the encoder. Each stage begins with a convolution layer as:
\begin{equation}
V^{(k)}= \begin{cases}
  H^{(k)} \star U^{(k)}& k\in\{0,4\},\\
  H^{(k)} \star_{\downarrow} U^{(k)}& \text{otherwise,}
\end{cases}
\label{eq:conv}
\end{equation}
where $\star$ and $\star_{\downarrow}$ are affine and down-sampling convolutions, respectively. Each convolution layer is followed by a GDN operation \cite{balle2016} defined as:
\begin{equation}
    w_{i}^{(k)}(m,n)=\frac{v_{i}^{(k)}(m,n)}{\sqrt{\beta^{(k)}_{i}+\sum_j{\gamma^{(k)}_{ij}\left(v_{j}^{(k)}(m,n)\right)^2}}},
\label{eq:gdn}
\end{equation}
where $i$ and $j$ run over channels and $(m,n)$ is the spatial location of a specific value of a tensor (e.g., $V^{(k)}$). Except for the first and last stages, a ResGDN transform is applied at the end of each stage:
\begin{equation}
U^{(k+1)} = \begin{cases}
  W^{(k)} & k=0,\\
  T(W^{(k)})+W^{(k)} & \text{otherwise,}
\end{cases}
\end{equation}
where $T$ is composed of two subsequent pairs of affine convolutions (Eq. \ref{eq:conv} with $k=0$), each is followed by a GDN operation (Eq. \ref{eq:gdn}).

\subsection{Stochastic Rounding-Based Quantization}
\label{sec:quantization}

The output of the last stage of the encoder, $W^{(4)}$, represents the code map $c$ with $\alpha_c$ channels. Each channel denoted by $c^{(i)}$ is then quantized to a discrete-valued vector using a stochastic rounding-based uniform scalar quantizer as: $c_q^{(i)}=Q(c^{(i)})$ where the function $Q$ is defined as in \cite{toderici2015variable,gupta2015deep,Raiko15}: $Q(c^{(i)}) = Round\left(\frac{c^{(i)}+\epsilon}{\Delta}\right)+z,$ where $\epsilon\in[-\frac{1}{2},\frac{1}{2}]$ is produced by a uniform random number generator. $\Delta$ and $z$ respectively represent the quantization step (scale) and the zero-point, which are defined as:
\begin{equation}
\begin{split}
    \Delta = \frac{max(c^{(i)})-min(c^{(i)})}{2^B-1},
z = \begin{cases}
  0 & \Bar{z} < 0,\\
2^B - 1 & \Bar{z} > 2^B - 1,\\
  \Bar{z} & \text{otherwise,}
\end{cases}
\end{split}
\label{eq:step_offset}
\end{equation}
where $\Bar{z}=\frac{-min(c^{(i)})}{\Delta}$, $B$ is the number of bits, and $min(c^{(i)})$ and $max(c^{(i)})$ are the input's min and max values over the $i$th channel, respectively. The zero-point $z$ is an integer ensuring that zero is quantized with no error, which avoids quantization error in common operations such as zero padding \cite{krishnamoorthi2018quantizing}. 

The stochastic rounding approach provides a better performance and convergence rate compared to round-to-nearest algorithm. Stochastic rounding is indeed an unbiased rounding scheme, which maintains a non-zero probability of small parameters. In other words, it possesses the desirable property that the expected rounding error is zero, i.e. $\mathbb{E}\left(Round(x)\right)=x$. As a consequence, the gradient information is preserved and the network is able to learn with low bits of precision without any significant loss in performance. 

Each of the quantized code map channels, denoted by $c^{(i)}_{q}$, is separately entropy-coded using the FLIF codec \cite{sneyers2016flif}, which is a state-of-the-art lossless image codec.

\subsection{Deep Decoder}

At the decoder side, the quantized code map is dequantized using the following function: $\hat{Q}(c^{(i)}_{q}) = \Delta.\left(c^{(i)}_{q}-z \right).$ Given the dequantized code map $\hat{c}^{(i)}_{q}=\hat{Q}(c^{(i)}_{q})$, the parametric decoder $f_{D}$ (with the parameter vector $\Psi$) reconstructs the image $x' \in \mathbb{R}^{h\times w\times 3}$ as follows: $x'=f_{D}(\hat{c}_q;\Psi)$. 
Similar to the encoder, the decoder network is composed of 5 stages in which all the operations are reversed. Each stage at the decoder network starts with an IGDN operation computed as follows:
\begin{equation}
    \hat{v}_{i}^{(k)}(m,n)={\hat{w}_{i}^{(k)}(m,n)}.{\sqrt{\hat{\beta}^{(k)}_{i}+\sum_j{\hat{\gamma}^{(k)}_{ij}\left(\hat{w}_{j}^{(k)}(m,n)\right)^2}}},
\label{eq:igdn}
\end{equation}
which is followed by a convolution layer defined as:
\begin{equation}
  \hat{U}^{(k)}= \begin{cases}
  \hat{H}^{(k)} \star \hat{V}^{(k)} & k\in\{0,4\},\\
  \hat{H}^{(k)} \star_{\uparrow} \hat{V}^{(k)} & \text{otherwise,}
\end{cases}
\label{eq:deconv}
\end{equation}
where $\star_{\uparrow}$ denotes transposed convolution used for upsampling the input tensor. As the reverse of the encoder, each convolution at the middle stages is followed by a ResIGDN transform defined as:
\begin{equation}
\hat{Z}^{(k)} = \hat{T}(\hat{U}^{(k)})+\hat{U}^{(k)}, k\in[1,3]
\end{equation}
where $\hat{T}$ consists of two subsequent pairs of an IGDN operation (Eq. \ref{eq:igdn}) followed by an affine convolution (Eq. \ref{eq:deconv} with $k=0$). The reconstructed image $x'$ is finally resulted from the output of the decoder represented by $\hat{U}^{(4)}$.


\begin{figure*}
\centering
\begin{minipage}[b]{0.45\linewidth}
 \centering
  \centerline{\includegraphics[width=\textwidth]{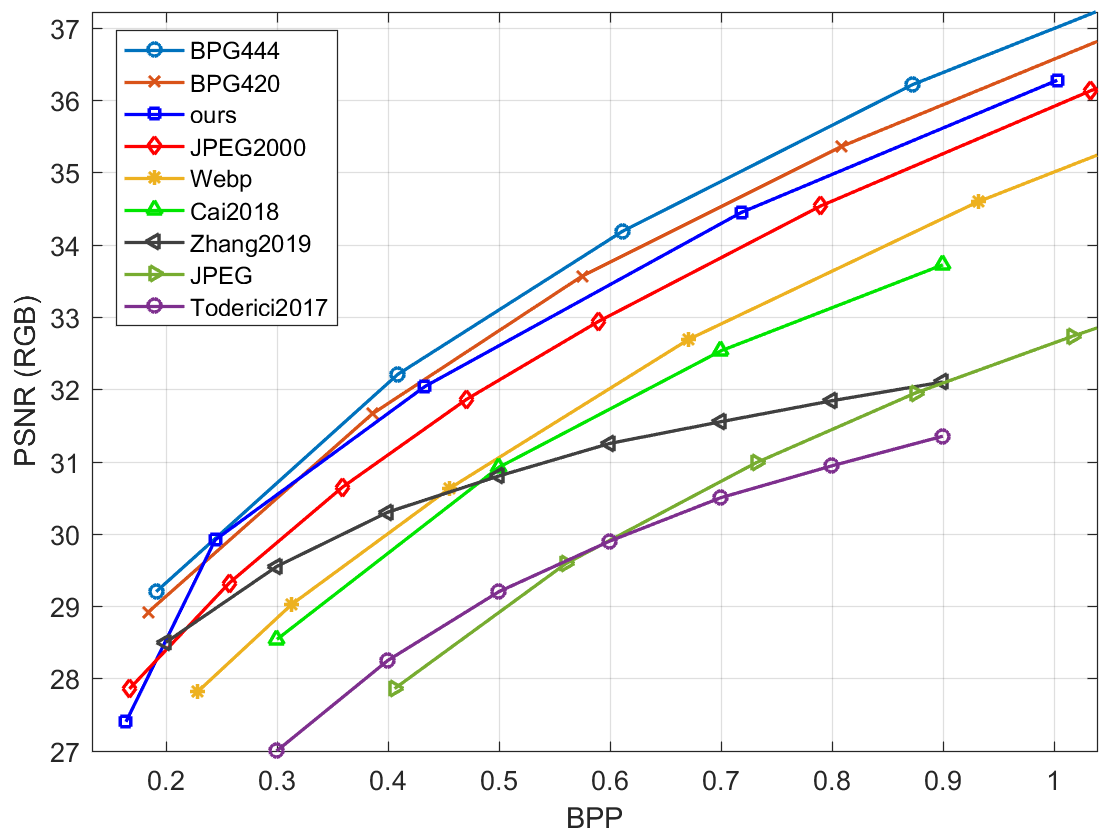}}
\end{minipage}
\begin{minipage}[b]{0.45\linewidth}
 \centering
  \centerline{\includegraphics[width=\textwidth]{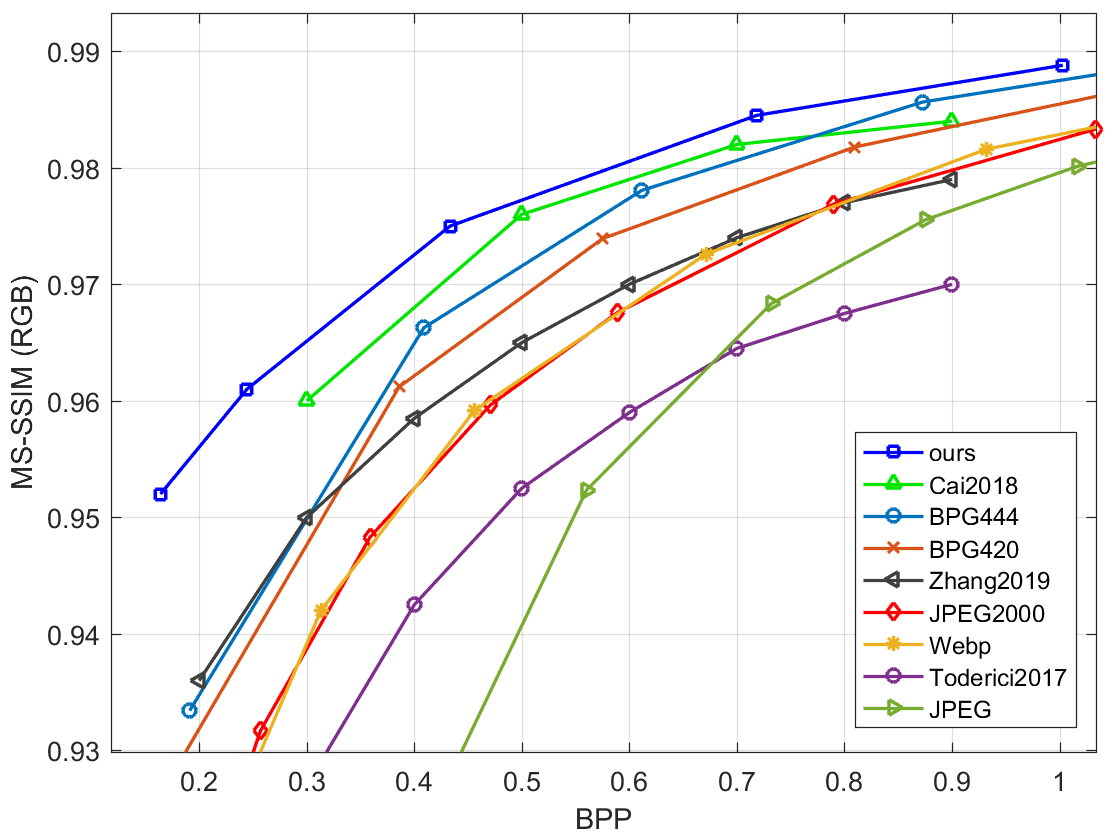}}
\end{minipage}
\caption{Comparison results of proposed variable-rate approach with state-of-the-art variable-rate methods on Kodak test set in terms of PSNR (left) and MS-SSIM (right) vs. bpp (bits/pixel).}
\vskip -3pt
\label{fig:scalable_results_Kodak}
\end{figure*}

\begin{figure*}
\centering
\begin{subfigure}[b]{.24\textwidth}
 \centering
  \centerline{\includegraphics[width=\textwidth]{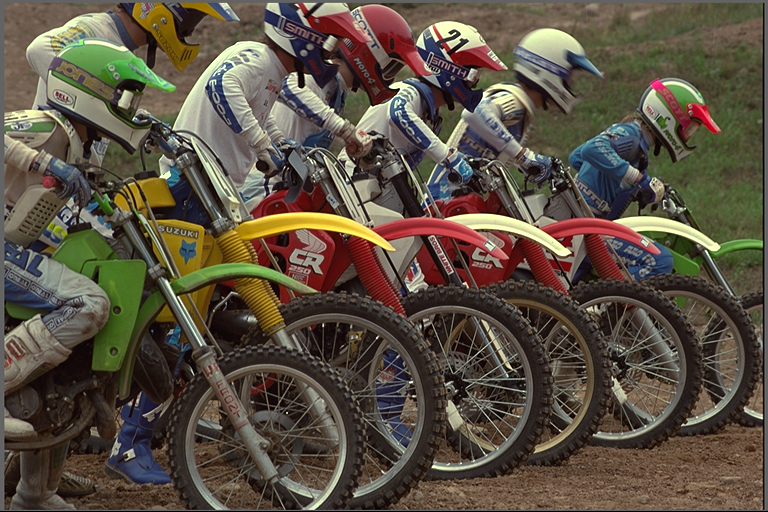}}
  \subcaption{\scriptsize Original}
\end{subfigure}
\begin{subfigure}[b]{.24\textwidth}
 \centering
  \centerline{\includegraphics[width=\textwidth]{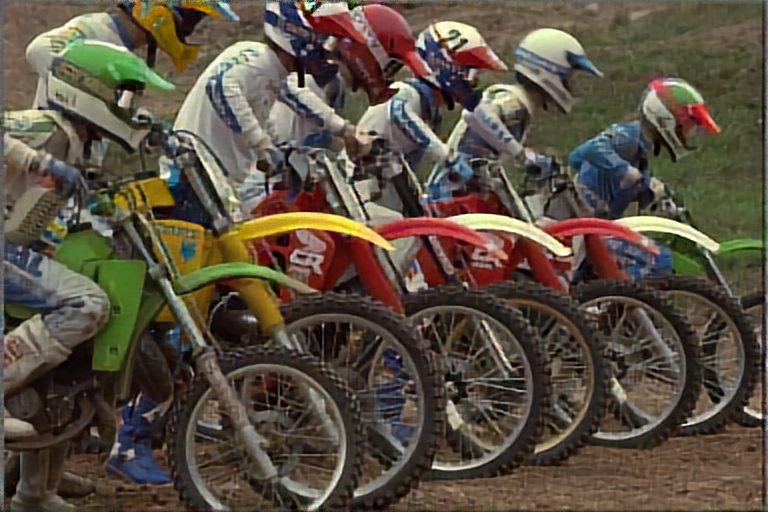}}
  \subcaption{\scriptsize ours (0.169bpp, 23.34dB, 0.947)}
\end{subfigure}
\begin{subfigure}[b]{.24\textwidth}
 \centering
  \centerline{\includegraphics[width=\textwidth]{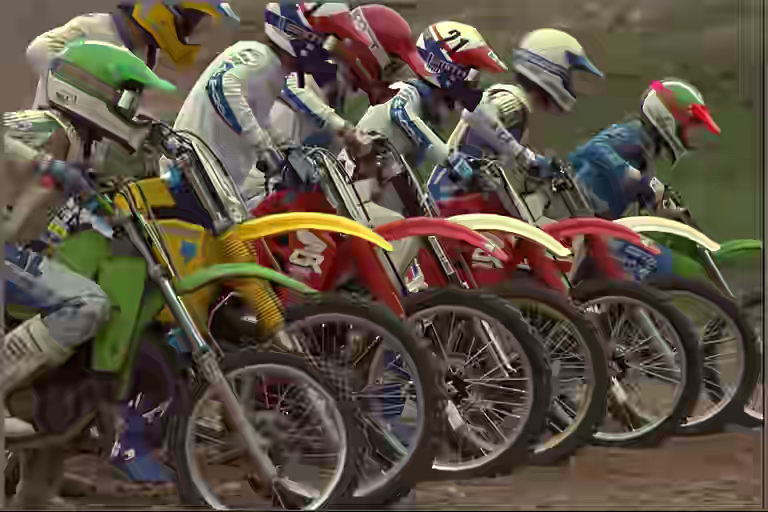}}
  \subcaption{\scriptsize BPG (0.177bpp, 23.66dB, 0.891)}
\end{subfigure}
\begin{subfigure}[b]{.24\textwidth}
 \centering
  \centerline{\includegraphics[width=\textwidth]{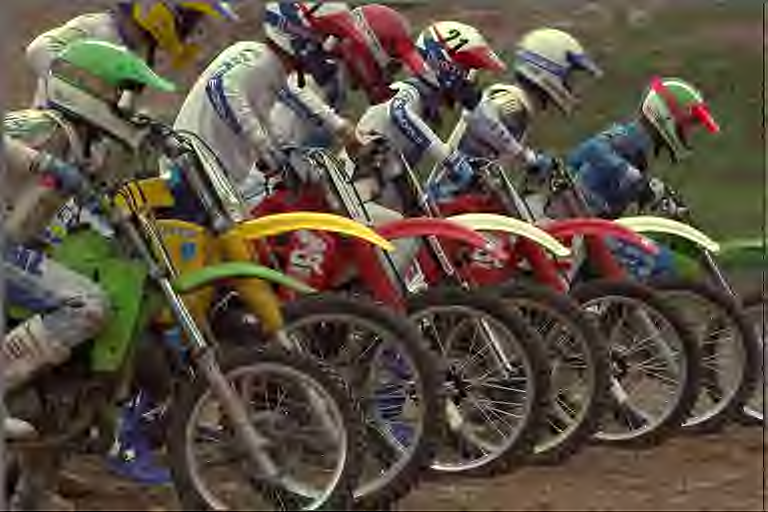}}
  \subcaption{\scriptsize JPEG2000 (0.175bpp, 22.47dB, 0.866)}
\end{subfigure}
\caption{Kodak visual example (bits/pixel, PSNR, MS-SSIM). \textit{BPG}: YUV4:4:4 format.}
\vskip -10pt
\label{fig:kodak_quan1}
\end{figure*}

\subsection{Residual Coding}

As an enhancement layer to the bit-stream, the residual $r$ between the input image $x$ and the deep decoder's output $x'$ is further encoded by the BPG codec, as in \cite{akbari2019dsslic}. To do this, the minimal and the maximal values of the residual image $r$ are first obtained, and the range between them is rescaled to [0,255], so that we can call the BPG codec directly to encode it as a regular 8-bit image. The minimal and maximal values are also sent to decoder for inverse scaling after BPG decoding.

\subsection{Objective Function and Optimization}

Our cost function is a combination of L2-norm loss denoted by $\mathcal{L}_{2}$, and MS-SSIM loss \cite{wang2004image} denoted by $\mathcal{L}_{MS}$ as follows:
\begin{equation}
\mathcal{L}(\Phi,\Psi) =  2\mathcal{L}_{2}+ \mathcal{L}_{MS},
\end{equation}
where $\Phi$ and $\Psi$ are the optimization parameter vectors of the deep encoder and decoder, respectively, each is defined as a full set of their parameters across all their 5 stages as: $\Phi=\{H^{(k)},\beta^{(k)},\gamma^{(k)}\}$ and $\Psi=\{\hat{H}^{(k)},\hat{\beta}^{(k)},\hat{\gamma}^{(k)}\}$ where $k=[0,4]$. In order to optimize the parameters such that our codec can operate at a variety of bit rates, we propose the following novel variable-rate objective functions for the $\mathcal{L}_{2}$ and $\mathcal{L}_{MS}$ losses:
\begin{equation}
\label{eq_obj}
\begin{split}
\mathcal{L}_{2}   &= \sum_{B\in {R}} {\lVert x-x'_B\rVert}_2,\\
 \mathcal{L}_{MS} &=-\sum_{B\in {R}} I_{M}(x,x'_B)\prod_{j=1}^M{ C_{j}(x,x'_B).S_{j}(x,x'_B)},
\end{split}
\end{equation}
where $x'_B$ denotes the reconstructed image with $B$-bit quantizer (Eq. \ref{eq:step_offset}), and $B$ can take all possible values in a set ${R}$. In this paper, ${R}=\{2,4,8\}$ is used for training variable-rate network model. 
The MS-SSIM metric use luminance $I$, contrast $C$, and structure $S$ to compare the pixels and their neighborhoods in $x$ and $x'$. Moreover, MS-SSIM operates at multiple scales where the images are iteratively downsampled by factors of  $2^{j}$, for $j \in[1,M]$. Note that other methods optimize for PSNR and MS-SSIM separately in order to get better performance in each of them. Our scheme jointly optimize for both of them. It will be shown later that we can still achieve satisfactory results in both metrics.

Our goal is to minimize the objective $\mathcal{L}(\Phi,\Psi)$ over the continuous parameters $\{\Phi, \Psi\}$. However, both depend on the quantized values of $c_q$ whose derivative is discontinuous, which makes the quantizer non-differentiable \cite{balle2016}. To overcome this issue, the fact that the exact derivatives of discrete variables are zero almost everywhere is considered, and the straight-through estimate (STE) approach in \cite{bengio2013estimating} is employed to approximate the differentiation through discrete variables in the backward pass.


\begin{figure*}
 \centering
\begin{minipage}[b]{0.45\linewidth}
 \centering
  \centerline{\includegraphics[width=\textwidth]{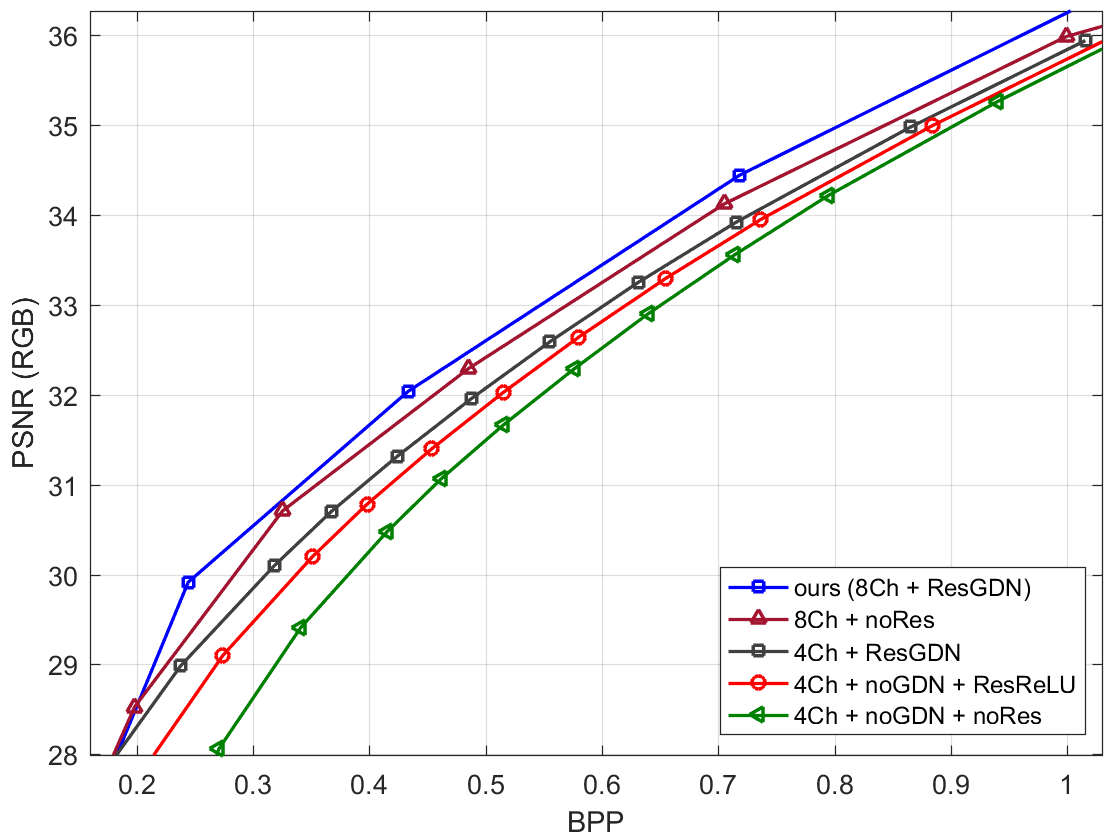}}
\end{minipage}
\begin{minipage}[b]{0.45\linewidth}
 \centering
  \centerline{\includegraphics[width=\textwidth]{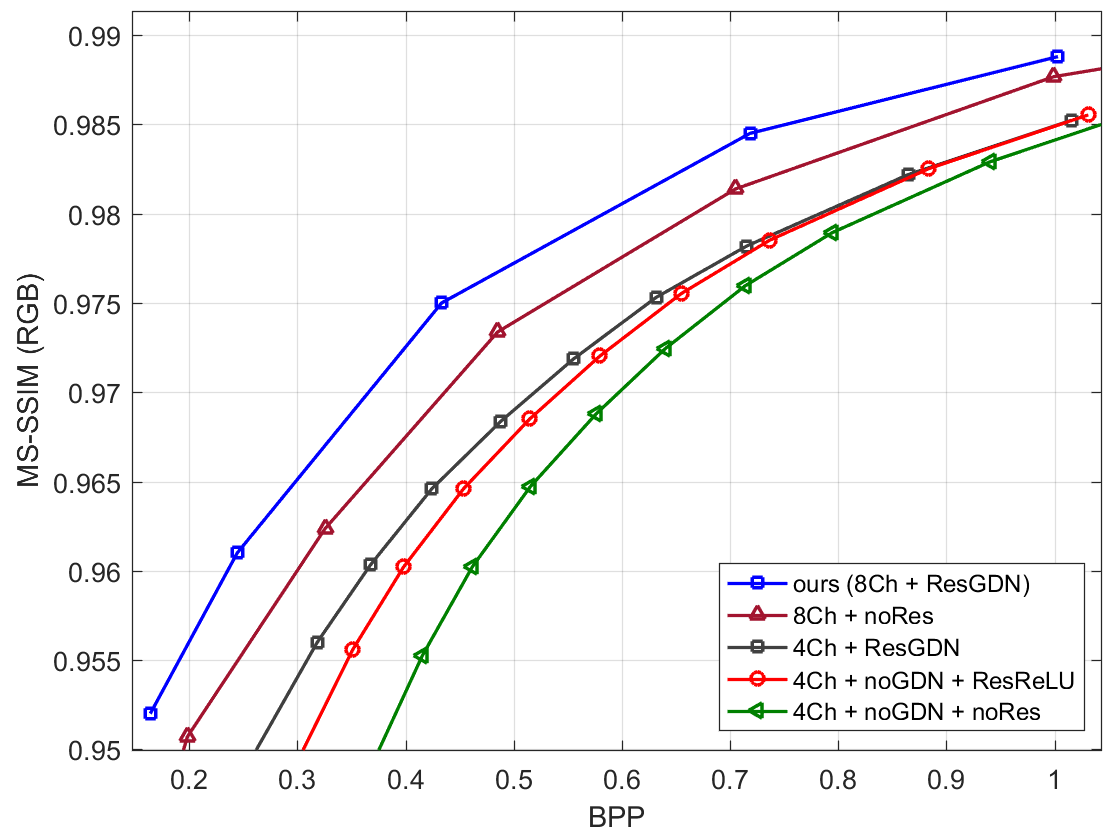}}
\end{minipage}
\caption{Ablation studies with different model configurations. \textbf{ours}: our variable-rate results. \textbf{\textit{n}Ch}: $n$ channels for the code map; \textbf{ResGDN}: ResGDN/IGDN transforms in the network architecture; \textbf{ResReLU}: conventional residual block with ReLU; \textbf{noRes}: neither ResGDN nor ResReLU is used; \textbf{noGDN}: GDN and IGDN layers in our main architecture replaced by ReLU.}
\vskip -10pt
\label{fig:ablation_results_Kodak}
\end{figure*}

\section{Experimental Results}

The ADE20K dataset \cite{zhou2017scene} was used for training the proposed model. The images with at least 512 pixels in height or width were used (9272 images in total). All images were rescaled to $h=256$ and $w=256$ to have a fixed size for training. We set the downsampling factor $\alpha=8$ and the channel size $\alpha_c=8$ to get the code map of size 32$\times$32$\times$8. The deep encoder and decoder models were jointly trained for 200 epochs with mini-batch stochastic gradient descent (SGD) and a mini-batch size of 16. The Adam solver with learning rate of 0.00002 was fixed for the first 100 epochs, and was gradually decreased to zero for the next 100 epochs. All the networks were trained in the RGB domain. The model was trained using 3 different bit rates, i.e., ${R}=\{2,4,8\}$ in Eq. \ref{eq_obj}. However, the trained model can operate at any bit rate in range $[1,8]$ at the test time. 

In this section, we compare the performance of the proposed scheme with standard codecs (including JPEG, JPEG2000, WebP, and the H.265/HEVC intra coding-based BPG codec \cite{bellard2017bpg}) and the state-of-the-art learning-based variable-rate image compression methods in Cai2018 \cite{cai2018efficient}, Zhang2019 \cite{toderici2017full}, and Toderici2017 \cite{zhang2018learned}, in which a single network was trained to generate multiple bit rates. We use both PSNR and MS-SSIM \cite{wang2003multiscale} as the evaluation metrics. 

The comparison results on the popular Kodak image set (averaged over 24 test images) are shown in Figure \ref{fig:scalable_results_Kodak}. Different points on the rate-distortion (R-D) curve of our variable-rate results are obtained from 5 different bit rates for the code maps in the base layer, i.e., $R = \{3,4,5,6,7\}$. The corresponding residual images $r$ in the enhancement layer are coded by BPG (YUV4:4:4 format) with quantizer parameters of $\{50, 40, 35, 30, 25\}$ respectively. As shown in Figure \ref{fig:scalable_results_Kodak}, our method outperforms the state-of-the-art learning-based variable-rate image compression models and JPEG2000 in terms of both PSNR and MS-SSIM. Our PSNR results are slightly lower than BPG (in both YUV4:2:0 and YUV4:4:4 formats), but we achieve better MS-SSIM, especially at low rates.

The BPG-based residual coding in our scheme is  exploited to avoid re-training the entire model for another bit rate and more importantly to boost the quality at high BPPs (bits/pixel). 
For the 5 points (low to high) in Figure \ref{fig:scalable_results_Kodak}, the percentage of BPPs used by residual image is \{2\%, 34\%, 52\%, 68\%, 76\%\}. This shows that as the BPP increases, the residual coding has more significant contribution to the R-D performance.




One visual example from the Kodak image set is given in Figures \ref{fig:kodak_quan1} in which our results 
are compared with JPEG2000 and BPG (YUV4:4:4 format). As seen in the example, our proposed method provides the highest visual quality compared to the others. JPEG2000 has poor performance due to the ringing artifacts at edges. The BPG result is smoother compared to JPEG2000, but the details and fine structures (e.g., the grass on the ground) are not preserved in many areas.

In order to evaluate the performance of different components of the proposed framework,  other ablation studies were performed. The results are shown in Figure \ref{fig:ablation_results_Kodak}.

\textbf{Code map channel size}: Figure \ref{fig:ablation_results_Kodak} shows the results with channel sizes of 4 and 8 (i.e., $\alpha_{c}\in\{4,8\}$). It can be seen that $\alpha_{c}=8$ has better results than $\alpha_{c}=4$. In general, we find that a larger code map channel size with smaller quantization bits provides a better performance, because deeper texture information of the input is preserved within the feature map. 



\textbf{GDN vs. ReLU}: In order to show the performance of the proposed GDN-based architecture, we compare it with a ReLU-based variant of our model, denoted as $noGDN$. In this model, all GDN and IGDN layers in the deep encoder and decoder are removed; instead, instance normalization followed by ReLU are added to the end of all convolution layers. The last GDN and IGDN layers in the encoder and decoder are replaced by a Tanh layer. As shown in Fig. \ref{fig:ablation_results_Kodak}, the models with GDN structure outperform the ones without GDN.


\textbf{Conventional vs. GDN/IGDN-based residual transforms}: In this scenario, the model composed of the proposed ResGDN/ResIGDN transforms (denoted by ResGDN in Figure \ref{fig:ablation_results_Kodak}) is compared with the conventional ReLU-based residual block (denoted by $ResReLU$) in which all the GDN/IGDN layers are repalced by ReLU. The results with  neither ResReLU nor ResGDN (yellow blocks in Figure \ref{fig:framework}), denoted by $noRes$, are also included. As demonstrated in Figure \ref{fig:ablation_results_Kodak}, ResGDN achieves better performance compared to the other scenarios.


In terms of complexity, the average processing time of the deep encoder, quantizer, and deep decoder on Kodak are 65ms, 2ms, and 51ms on a TITAN X Pascal GPU, respectively.


\section{Conclusion}

In this paper, we propose a new variable-rate image compression framework, by applying more GDN layers and incorporating the shortcut connection in the ResNet. We also use a stochastic rounding-based scalable quantization. To further improve the performance, the residual between the input and the reconstructed image from the decoder network is encoded by BPG as an enhancement layer. A novel variable-rate objective function is proposed. Experimental results show that our variable-rate model can outperform all standard codecs including BPG, as well as state-of-the-art learning-based variable-rate methods. A future topic is the rate allocation optimization between the base layer and the enhancement layer.

\bibliographystyle{IEEEbib}
\bibliography{refs}

\end{document}